\documentclass[aps,prb,twocolumn,groupedaddress,floatfix,amsmath,amssymb,superscriptaddress]{revtex4-2}
\usepackage{amsmath}
\usepackage{amssymb}
\usepackage{braket}
\usepackage{graphicx}
\usepackage[table]{xcolor}
\usepackage{verbatim}
\usepackage{float}
\usepackage{color}
\usepackage{siunitx}
\usepackage{gensymb}
\usepackage{bm}
\usepackage{xcolor}
\usepackage{multirow}
\usepackage{footnote}
\usepackage{bbm}

\usepackage{array}
\newcolumntype{P}[1]{>{\centering\arraybackslash}p{#1}}
\newcolumntype{Q}[1]{>{\raggedleft\arraybackslash}p{#1}}
\newcolumntype{R}[1]{>{\raggedright\arraybackslash}p{#1}}

\newcommand{\rr}{\mathbf{r}}

\newcommand{\qq}{\mathbf{q}}

\newcommand{\GG}{\mathbf{G}}

\newcommand{\bvec}[1]{\mathbf{\boldsymbol{#1}}}

\begin{document}

\title{Interlayer dislocations in multilayer and bulk MoS${}_2$}

\author{Isaac Soltero}
\email{isaac.solteroochoa@manchester.ac.uk}
\affiliation{Department of Physics and Astronomy, University of Manchester. Oxford Road, Manchester, M13 9PL, United Kingdom}
\affiliation{National Graphene Institute, University of Manchester. Booth St.\ E., Manchester, M13 9PL, United Kingdom}
\author{Vladimir I. Fal'ko}
\email{vladimir.falko@manchester.ac.uk}
\affiliation{Department of Physics and Astronomy, University of Manchester. Oxford Road, Manchester, M13 9PL, United Kingdom}
\affiliation{National Graphene Institute, University of Manchester. Booth St.\ E., Manchester, M13 9PL, United Kingdom}

\begin{abstract}
    Dislocations in van der Waals materials are linear defects confined to the interfaces between consecutive stoichiometric monolayers of a bulk layered crystal. Here, we present a mesoscale model for the description of interlayer dislocations in thin films of transition metal dichalcogenides. Taking 2H-MoS${}_2$ as a representative material, we compute the dependence of the dislocation energy on the film thickness, from few-layer MoS$_2$ to the bulk crystal, and analyse the strain field in the layers surrounding a dislocation. We also analyse the influence of strain field on the band edge profiles for electrons and holes, and conclude that the resulting energy profiles are incapable of localising charge carriers, in particular at room temperature.
\end{abstract}

\maketitle

\section{Introduction}

Dislocation lines in solids are fundamental topological defects \cite{anderson2017theory,landau2012theory,kosevich2006crystal}. In semiconductor technologies, dislocations represent a problem for device performance since these entail a deterioration of electronic and optical properties. In particular, in technologically relevant silicon and germanium, dangling bonds along dislocations lead to electronic states in the band gap capable of trapping electrons and holes \cite{grazhulis1977investigation,eremenko1977dependence,kusanagi1992difference}, resulting in leaking semiconductor structures and charge inhomogeneity.

Dislocations also appear in bulk layered materials with van der Waals (vdW) interlayer coupling, such as hexagonal transition metal dichalcogenides (TMDs). In such materials, one can distinguish two types of dislocations: (i) those with axes piercing the layers and (ii) and in-plane screw dislocations with axes oriented along the layers, characterised by substantial interlayer shear deformations. While the former type of dislocations are energetically expensive and rarely appear in high quality grown vdW materials, the latter type is quite common in various exfoliated structures \cite{zhang2014three,fan2018controllable,butz2014dislocations}. For a vdW bilayer, an interlayer dislocation is nothing but an incommensurate boundary (``domain wall'') separating two commensurate stacking regions  \cite{alden2013strain,butz2014dislocations,lebedeva2016dislocations,enaldiev2020stacking,enaldiev2024dislocations}. To mention, networks of such dislocations naturally appear at interfaces between two marginally twisted bilayers \cite{carr2018relaxation,naik2018ultraflatbands,yoo2019atomic,weston2020atomic,mcgilly2020visualization,rosenberger2020twist}, or multilayer films of vdW crystals \cite{cook2023moire,halbertal2023multilayered,zhu2020modeling,liang2020effect,craig2024local,nakatsuji2023multiscale,ceferino2024pseudomagnetic}.

Here, we assess the ability of an in-plane screw dislocation in a vdW semiconductors to trap charge carriers, using 2H-MoS${}_2$ as a representative example. For this, we analyse the deformation field associated with a single dislocation in the middle of a thin film of a TMD, including bilayers, tetralayers, hexalayers, and bulk crystals. We take into account piezoelectric properties of TMD monolayers, caused by lack of inversion symmetry of their unit cell, and calculate the piezocharge distribution in each layer near a dislocation in a MoS${}_2$ film. For convenience, we consider films with $2N$ layers with the dislocation plane at the interface of the $N$th and $(N+1)$th monolayers. 

After combining piezo-potentials created by piezocharges with the influence of interlayer shear on conduction/valence band energies, we obtain maps of band edge profiles for electrons and holes. The analysis of such profiles performed in this paper indicates that a screw dislocation in a film with up to 20 layers is not able to trap carriers, neither electrons nor holes. Moreover, we find that, while a screw dislocation in a 3D bulk of 2H-MoS${}_2$ can weakly bind electrons (but not holes), the estimated sub-meV binding energy indicates that it would not be charged at the room temperature. This result suggests that perfect screw dislocations in TMD films are benign defects in terms of their effect on performance of TMD-based field effect transistors.

\begin{figure}[h!]
    \centering
    \includegraphics[width=\columnwidth]{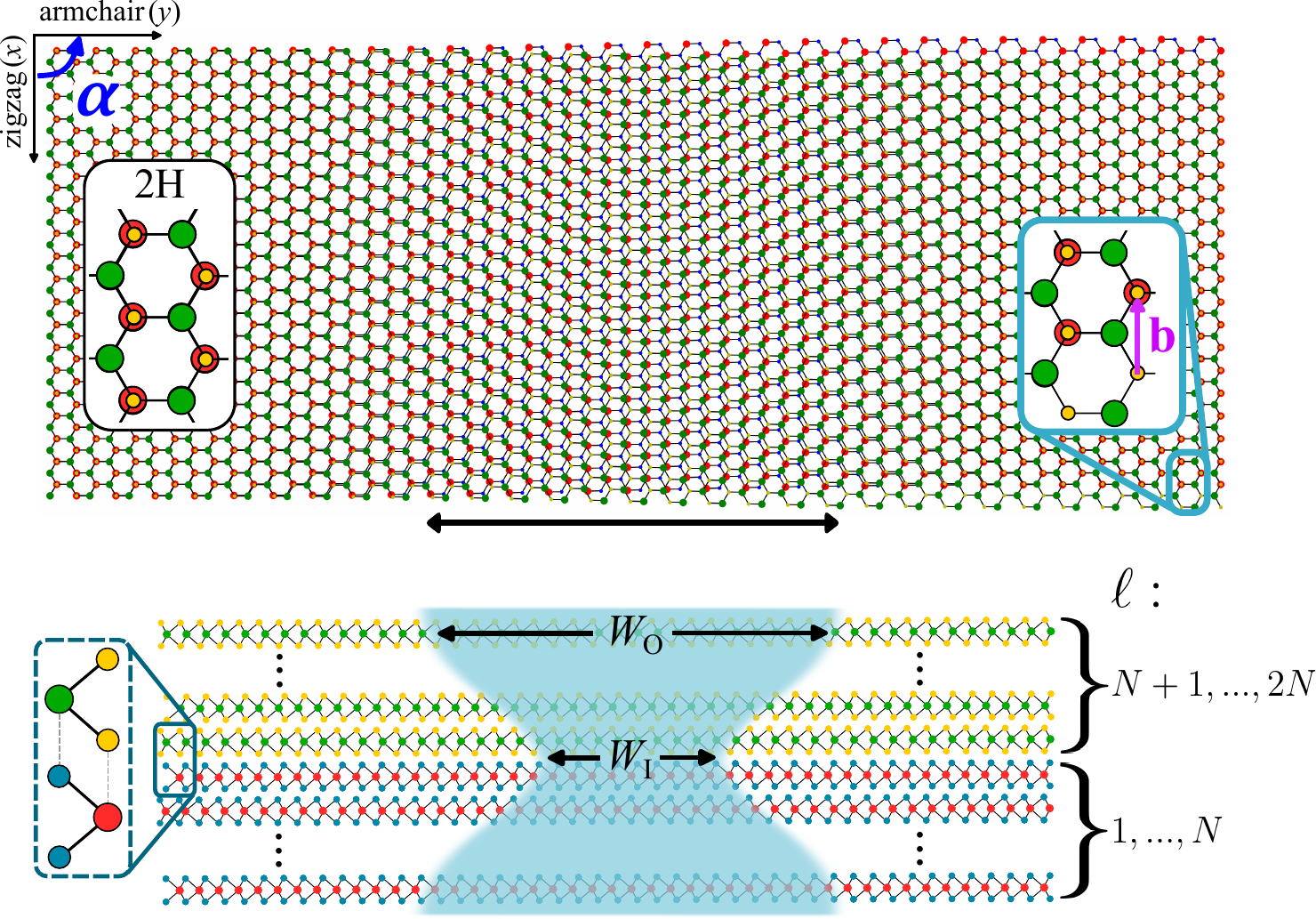}
    \caption{Interlayer 2H/2H screw dislocation in a MoS${}_2$ bilayer (top panel). Insets show the top view of the stacking in commensurate regions, where the Burgers vector $\bvec{b}$ is indicated. The dislocation axis $\alpha$ is depicted with a blue arrow (in this case $\alpha = 90^{\circ}$, $\xi=y$). Side view of a dislocation in a $2N$ layer structure (bottom panel), with inset outlining the alignment between chalcogens and metals between neighbouring layers. Dislocation plane is located between layers with indexes $\ell = N$ and $N+1$. Blue region indicates the width profile of the dislocation, acquiring a minimum of $W_{\rm I}$ in the interface layers, and a maximum of $W_{\rm O}$ in the outer layers.}
    \label{fig:Stacking}
\end{figure}

The rest of the paper is organised as follows. In Section \ref{Sec:Model}, we discuss our theoretical model of interlayer dislocations in multilayer TMDs. In Sec. \ref{SubSec:DislocationStrc} we analyse the energetics and structural properties of dislocations and their dependence on the number of layers in 2H-MoS${}_2$. The effect of shear-induced piezoelectric charges on electron/hole band edges across dislocations in few-layer films is discussed in Sec. \ref{SubSec:Piezo}. In Sec. \ref{Sec:Bulk} we study the piezo-potential landscape and confinement of electrons in bulk MoS${}_2$. Our concluding remarks are presented in Sec. \ref{Sec:Conclusions}.

\section{Model}\label{Sec:Model}

Interlayer dislocations in 2H TMD crystals, where consecutive layers have antiparallel orientation of their unit cells, are formed between stacking regions with metals over chalcogens and chalcogens over metals simultaneously (from now on called 2H stacking). The energy and interlayer distance variation due to the smooth change of stacking across 2H/2H dislocations is taken into account via an analytically interpolated adhesion energy density between neighbouring layers in the structure \cite{enaldiev2020stacking},
\begin{equation}
\begin{split}
    \mathcal{W}(\rr_0,d) = \sum_{n=1}^{3}\bigg[& -\frac{C_{n}}{d^{4n}} + A_{1}e^{-\sqrt{G^{2}+\varrho^{-2}}d}\cos\big(\GG_{n}\cdot\rr_0\big)\\
    & + A_{2}e^{-Gd}\sin\big(\GG_{n}\cdot\rr_0\big) \bigg].
\end{split}
\end{equation}
Here, $d$ is the distance between layers, $\rr_{0}$ is an in-plane offset vector between nearest chalcogens of each layer, and $\GG_{n}$ (with $G\equiv |\GG_{n}|$) are reciprocal lattice vectors of the monolayer crystals. Parameters $C_{n}$, $A_{1,2}$ and $\varrho$ are established from DFT modelling and are shown in Table \ref{tab:Parameters} for MoS${}_2$.

Dislocations in a $2N$ layer system are described by elastic deformations in each layer $\ell$. To describe the on-layer displacement fields, we use a coordinate system where the orientation of the dislocation axis is at an angle $\alpha$ accounted from the zigzag direction of the TMD lattice, and the in-plane and out-of-plane deformation fields, $\bvec{u}^{(\ell)}$ and $\zeta^{(\ell)}$, depend on the distance $\xi$ from the dislocation axis. At long distances, we impose the boundary conditions,
\begin{subequations}
\begin{equation}
    \bvec{u}^{(\ell)}(-\infty)=\bvec{0},
\end{equation}
\begin{equation}
    \bvec{u}^{(\ell)}(\infty) =
    \begin{cases}
        (a/2)\hat{\boldsymbol{\mathrm{x}}} & \ell\leq N \\
        -(a/2)\hat{\boldsymbol{\mathrm{x}}} & \ell\geq N + 1
    \end{cases},
\end{equation}
\begin{equation}
    \zeta^{(\ell)}(-\infty) = \zeta^{(\ell)}(\infty) = 0.
\end{equation}
\end{subequations}
This defines a dislocation with Burgers vector $\bvec{b}=\bvec{u}^{(\ell)}(\infty) - \bvec{u}^{(\ell')}(\infty) = -a\hat{\boldsymbol{\mathrm{x}}}$ (for any $\ell\geq N+1$ and $\ell'\leq N$), where $a=0.316$ nm is the monolayer lattice constant \cite{wilson1969transition}. The variation of in-plane offset vector and the interlayer distance for every pair of neighbouring layers in the structure reads as
\begin{subequations}
\begin{equation}
\begin{split}
    \rr_0^{(\ell,\ell+1)}(\xi) =&\, \rr_0^{\rm (2H)} + (-1)^{\ell}\big[ u_{x}^{(\ell+1)}(\xi) - u_{x}^{(\ell)}(\xi) \big] \hat{\boldsymbol{\mathrm{x}}} \\
    & + \big[ u_{y}^{(\ell+1)}(\xi) - u_{y}^{(\ell)}(\xi) \big] \hat{\boldsymbol{\mathrm{y}}},
\end{split}
\end{equation}
\begin{equation}
    d^{(\ell,\ell+1)}(\xi) = d^{\rm (2H)} + \zeta^{(\ell+1)}(\xi) - \zeta^{(\ell)}(\xi).
\end{equation}
\end{subequations}
Here, $\rr_{0}^{\rm (2H)} = (a/2)(\hat{\boldsymbol{\mathrm{x}}} + \hat{\boldsymbol{\mathrm{y}}}/\sqrt{3})$ and $d^{\rm (2H)} = 0.625$ nm are the offset vector and interlayer distance for perfect 2H stacking, respectively, and the factor $(-1)^{\ell}$ accounts for the opposite orientation of the unit cells between consecutive layers.

\begin{table}[t!]
    \caption{Adhesion energy density parameters \cite{enaldiev2020stacking}, elastic parameters \cite{iguiniz2019revisiting,androulidakis2018tailoring} and piezocoefficient \cite{zhu2015observation} for 2H-MoS${}_2$.}
    \centering
    \begin{tabular}{c c c c c c}
    \hline
    \hline
       $C_{1}$ & $C_{2}$ & $C_{3}$  & $A_{1}$ & $A_{2}$ & $\varrho$ \\
       ${\rm eV}\cdot{\rm nm}^{2}$ & ${\rm eV}\cdot{\rm nm}^{2}$ & ${\rm eV}\cdot{\rm nm}^{2}$ & ${\rm eV}/{\rm nm}^{2}$ & ${\rm eV}/{\rm nm}^{2}$ & nm \\
       0.134661  & 0.161589 & -0.0209218 & 71928800 & 56411 & 0.0496 \\
    \hline 
    $Y$ & $\sigma$ & $\lambda$ & $\mu$ & $\kappa$ & $|e_{11}|$ \\
    GPa & -- & eV/${\rm nm}^2$ & ${\rm eV}/{\rm nm}^{2}$ & eV & ${\rm C}/m$\\
    277 & 0.27 & 520.1 & 443.0 & 37.9 & 2.9$\times 10^{-10}$ \\
    \hline
    \hline
    \end{tabular}
    \label{tab:Parameters}
\end{table}

The structure across the dislocation is found considering the energy density functional
\cite{landau2012theory,carr2018relaxation,enaldiev2020stacking}
\begin{equation}\label{EnergFun}
\begin{split}
    \mathcal{E} =&  \int_{-\infty}^{\infty} d\xi \Bigg[ \sum_{\ell=1}^{2N-1} \mathcal{W}\big(\rr_0^{(\ell,\ell+1)},d^{(\ell,\ell+1)}\big)+ \sum_{\ell = 1}^{2N} \big( \mathcal{U}_{\ell} + \mathcal{V}_{\ell}  \big)\Bigg],
\end{split}
\end{equation}
where the second term accounts for the energy cost of in-plane strain  and bending,
\begin{subequations}
\begin{equation}
    \mathcal{U}_{\ell} = \frac{\lambda}{2}\big(u_{ii}^{(\ell)}\big)^{2} + \mu u_{ij}^{(\ell)}u_{ji}^{(\ell)};
\end{equation}
\begin{equation}
\begin{split}
    \mathcal{V}_{\ell} =&\frac{\kappa}{2}\big( \partial_{x}^{2}\zeta^{(\ell)}+\partial_{y}^{2}\zeta^{(\ell)} \big)^{2}\\
    & +\frac{\kappa'}{2}\Big\{ \big(\partial_{x}\partial_{y}\zeta^{(\ell)}\big)^{2} - \big(\partial_{x}^{2}\zeta^{(\ell)}\big)\big(\partial_{y}^{2}\zeta^{(\ell)}\big) \Big\}.
\end{split}
\end{equation}
\end{subequations}
Here, $u_{ij}^{(\ell)}=(\partial_{i}u_{j}^{(\ell)} + \partial_{j}u_{i}^{(\ell)})/2$ is an in-plane
strain tensor for each layer, $\lambda$ is the first Lam{\'e} coefficient, $\mu$ is the shear modulus, and $\kappa$ the flexural rigidity. This parameters are expressed in terms of the bulk Young modulus, $Y$, and Poisson ratio, $\sigma$, as
\begin{subequations}
\begin{equation*}
    \lambda=\frac{Yd^{\rm (2H)}\sigma}{(1+\sigma)(1-2\sigma)}, \quad \mu = \frac{Yd^{\rm (2H)}}{2(1+\sigma)}, \quad \kappa = \frac{Y \big(d^{\rm (2H)}\big)^{3}}{12(1-\sigma^{2})},
\end{equation*}
\end{subequations}
and we have additionally defined $\kappa'\equiv 2(1-\sigma)\kappa$. 

The optimal configuration for the deformation fields is found by minimising the energy density with respect to all in- and out-of-plane configurations, i.e.  $u_{x}^{(\ell)}$, $u_{y}^{(\ell)}$ and $\zeta^{(\ell)}$ in every layer $\ell$. As integral in Eq. \eqref{EnergFun} diverges due to the contribution from bulk domains, we define, and then minimise, the dislocation energy per unit of length,
\begin{equation}\label{EnergDens}
    \mathcal{E}_{l} \equiv \mathcal{E} -(2N-1) \int_{-\infty}^{\infty}d\xi\, \mathcal{W}\big(\rr_0^{\rm (2H)},d^{\rm (2H)}\big),
\end{equation}
where the second term represents the energy of a structure with uniform 2H stacking. The dislocation strain field and local energy density were calculated on a mesh for $\xi$ coordinate with a step of $0.2$ nm and the length of integration interval was increased until achieving convergence. Miminisation was implemented using the limited-memory Broyden–Fletcher–Goldfarb–Shanno algorithm (L-BFGS), with a tolerance of $10^{-8}$ eV/nm. The obtained numerical results are described in detail in the next sections.

\section{Dislocations in few-layer M\lowercase{o}S${}_2$}\label{Sec:DislFL}

\subsection{Dislocation structural properties}\label{SubSec:DislocationStrc}

\begin{figure}
    \centering
    \includegraphics[width=0.9\columnwidth]{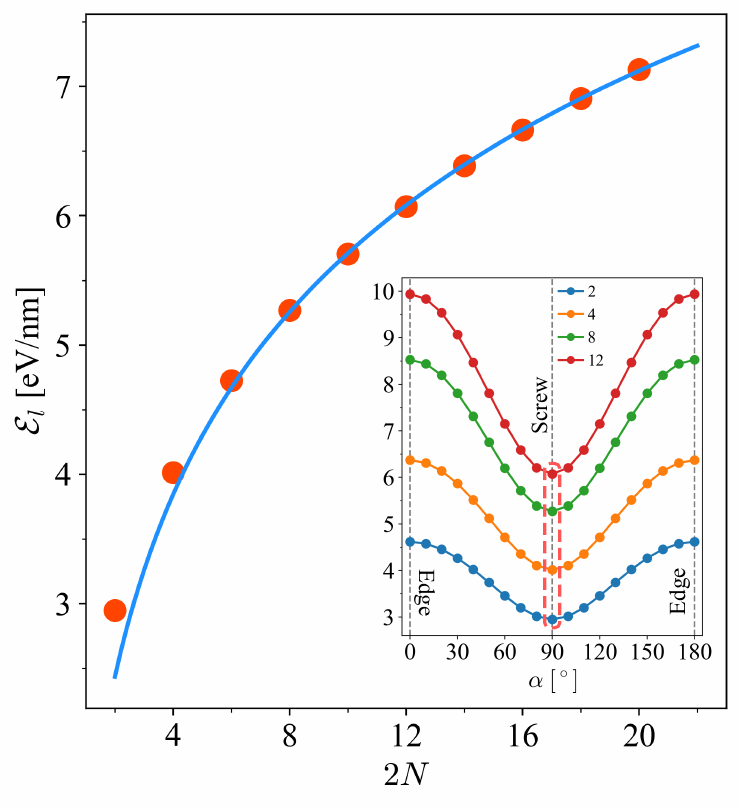}
    \caption{Dislocation energy per unit of length (defined by Eq. \eqref{EnergDens}) as a function of the number of layers for a screw dislocation in 2H-MoS${}_2$. Blue curve represents the logarithmic fit of the data for $2N$ from 6 to 20. Inset shows the dependence of the dislocation energy density with respect to the angle $\alpha$ between the zigzag direction and the dislocation axis for $2N=2,4,8$ and 12. $\alpha=0^{\circ}$ corresponds to a dislocation along the armchair direction (edge dislocation), whereas $\alpha=90^{\circ}$ is for a dislocation along the zigzag direction (screw dislocation). The later case is highlighted with a dashed box as the most energetically favourable configuration.}
    \label{fig:DW_Energy}
\end{figure}

The most favourable orientation of a dislocation in few-layer TMDs with respect to the constituent monolayers' crystallographic axis is found minimising \eqref{EnergDens} for a range of values $0^{\circ}\leq\alpha\leq 180^{\circ}$. As shown in the inset of Fig. \ref{fig:DW_Energy}, the highest energy configuration occurs for $\alpha=0^{\circ}$, which corresponds to a dislocation along the armchair direction (edge dislocation). The minimal energy density is found for a dislocation with its axis oriented along the zigzag direction ($\alpha=90^{\circ}$), which corresponds to a perfect screw dislocation: in this case the coordinate $\xi$ is counted along armchair axis ($y$), see Fig. \ref{fig:Stacking}(a). Below we analyse the dependence of such dislocation energy and deformation fields in MoS${}_2$ films with different thicknesses ($2\leq 2N \leq 20$).

The evolution of the screw dislocation energy density when increasing the number of layers (Fig. \ref{fig:DW_Energy}) shows a logarithmic divergence, which is captured by the fitting
\begin{equation}\label{LogFit}
    \mathcal{E}_{l}(N) \approx \beta \ln( 4 N), \quad \beta = 2.0 \, {\rm eV/nm},
\end{equation}
in agreement with the classical result that energy density grows logarithmically with the size of the crystal \cite{anderson2017theory}.

\begin{figure}
    \centering
    \includegraphics[width=0.9\columnwidth]{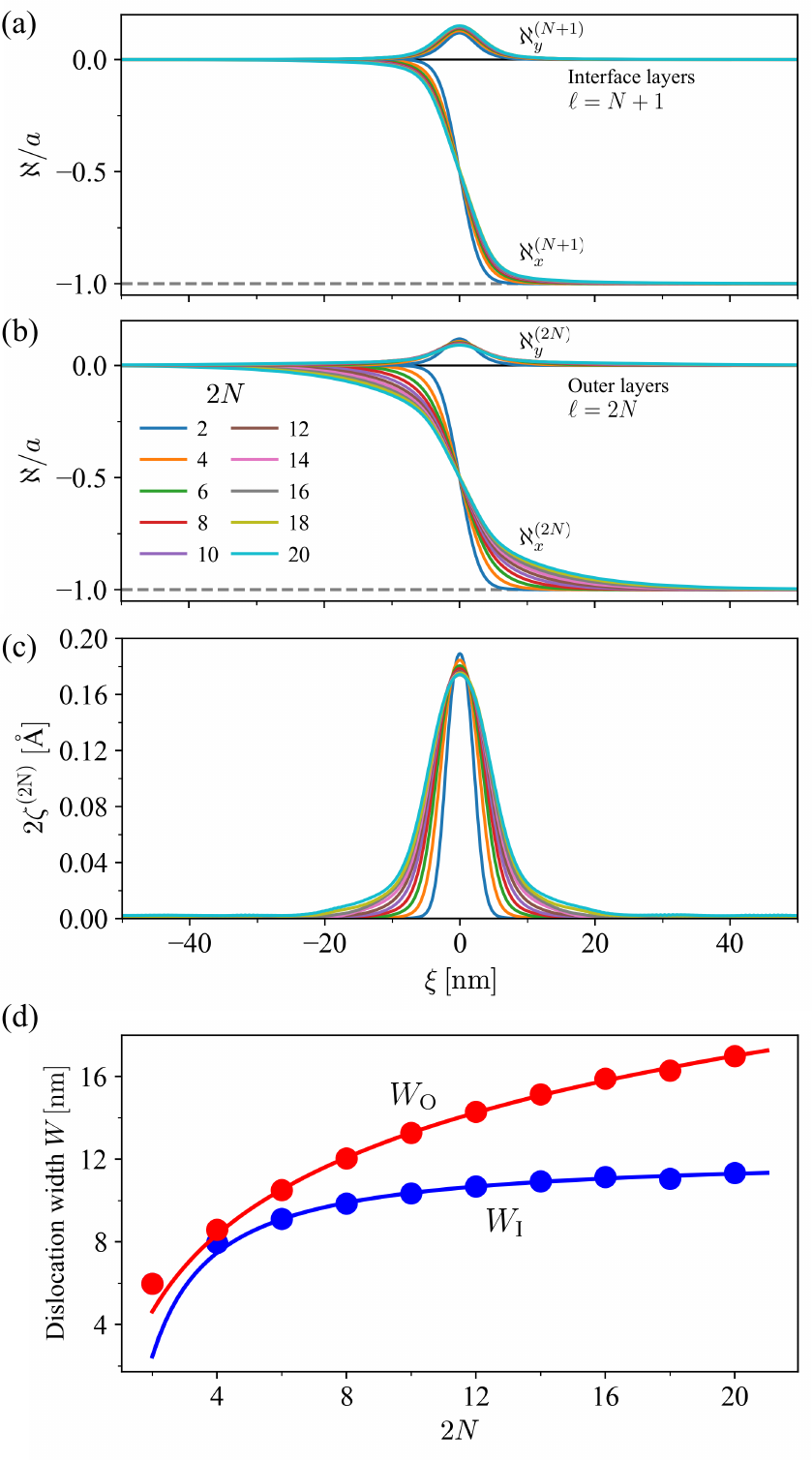}
    \caption{Displacement offset profiles $\aleph$ for (a) interface and (b) outer layers, and (c) variation of the thickness of the film across a screw dislocation in 2H-MoS${}_2$ crystals for different number of layers ($2N$). (d) Dislocation width in the interface (blue, $\ell = N,N+1$) and the outer layers (red, $\ell = 1,2N$) as a function of the number of layers. Circles correspond to the calculated values, while solid lines represent the fittings \eqref{FitInter} and \eqref{FitOut}.}
    \label{fig:DW_profiles}
\end{figure}

To image the profile of the dislocation, we use the interlayer offset defined as $\aleph_{x/y}^{(\ell)}(\xi)=u_{x/y}^{(\ell)}(\xi) - u_{x/y}^{(N)}(\xi)$, where $\ell\geq N + 1$, which is plotted in Fig. \ref{fig:DW_profiles}(a). Here, $\ell=N+1$ is the interface layer, and $\ell=2N$ is the outer layer. The presented profiles show how deformations spread out away from the center of the dislocation, which we use to estimate the dislocation width in each layer (using a linear fitting of $\aleph_{x}(\xi)$ around $\xi=0$ and taking the intersection with $\aleph=0$) shown in Fig. \ref{fig:DW_profiles}(d). To mention, the numerically found width of the deformation field distribution at the dislocation plane depends on the thickness of the film as
\begin{equation}\label{FitInter}
\begin{split}
    W_{\rm I}(N) \approx&\, \delta - \frac{\nu}{2N}, \quad \delta = 12.2 \, {\rm nm}, \,\, \nu=20.1\, {\rm nm},
\end{split}
\end{equation}
saturating at $W_{\rm I}(N\gg 2)\approx 12.2$ nm in a bulk material, which is twice the dislocation width in a bilayer. Also, the spread of deformation fields in the outer layer grows logarithmically with the film thickness, as
\begin{equation}\label{FitOut}
    W_{\rm O}(N) \approx \eta \ln(2.4\, N), \quad \eta = 5.4\, {\rm nm}.
\end{equation}
In Fig. \ref{fig:DW_profiles}(c) we show the variation of the thickness of the film, $2\zeta^{(2N)}$, across the dislocation profile, determined using the out-of-plane deformations of the outermost layer. From the dependence of the film thickness we conclude that the maximal `swelling' of the film is mostly determined by the out-of-plane deformations of the layers $\ell=N$ and $N+1$ interfacing each other at the dislocation plane, whereas the width of the swelled interval increases with the number of layers, as described by Eq. \eqref{FitOut}.

\subsection{Piezoelectric potentials for electrons and holes}\label{SubSec:Piezo}

The unit cell of TMD monolayers lacks inversion symmetry, so that in inhomogeneous shear strain in each layer of the film produces a piezoelectric charge density \cite{enaldiev2020stacking},  $\rho^{(\ell)} = e_{11}^{(\ell)}[2\partial_{x}u_{xy}^{(\ell)} + \partial_{y}(u_{xx}^{(\ell)} - u_{yy}^{(\ell)})]$, where the signs of piezocoefficients alternate from layer to layer, $e_{11}^{(\ell)}= (-1)^{\ell}e_{11}$, due to the inverted orientations of the unit cells in the consecutive monolayers of 2H crystals. The projection of this expression to the armchair direction ($\xi \equiv y$) reads as
\begin{equation}
\begin{split}
    \rho^{(\ell)}(\xi) = -e_{11}^{(\ell)} \partial_{\xi}^{2}u_{y}^{(\ell)}(\xi),
\end{split}
\end{equation}
where strain tensors are such that $u_{ij}^{(\ell)}=-u_{ij}^{(2N+1-\ell)}$ for $\ell\leq N$. All these  result in the alternating signs of charge density profiles between neighbouring layers, except those two at the dislocation plane ($N$ vs $N+1$), and $\rho^{(\ell)}=\rho^{(2N+1-\ell)}$.

\begin{figure}[t]
    \centering
    \includegraphics[width=1\columnwidth]{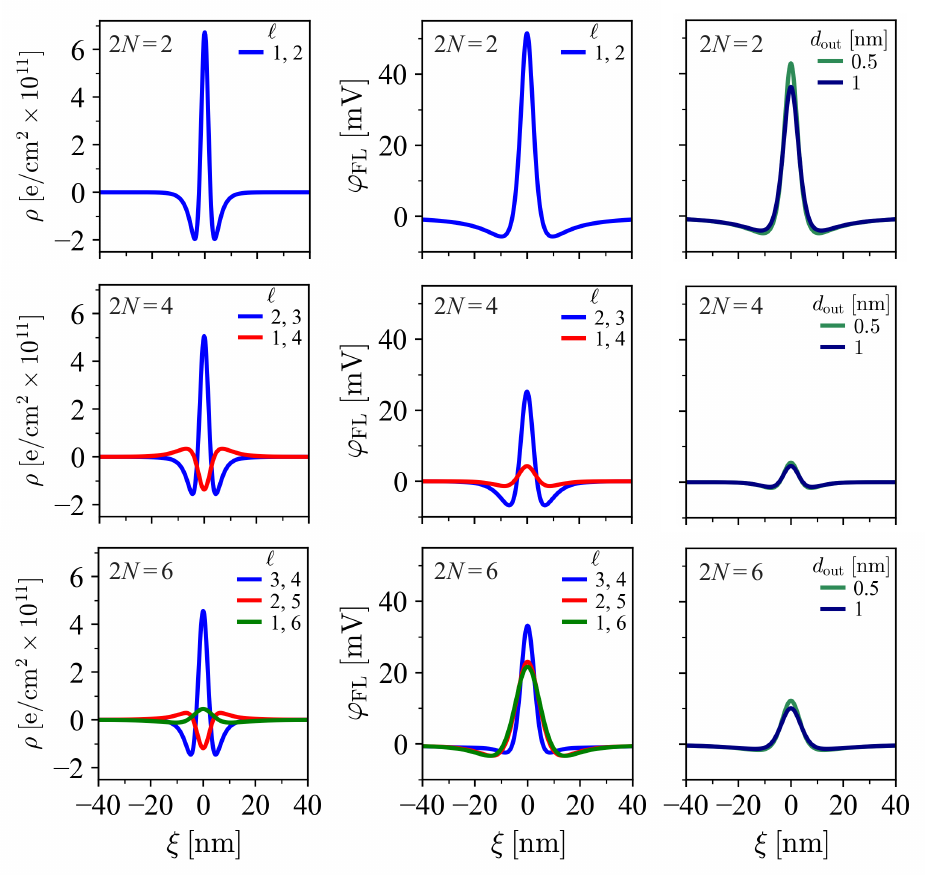}
    \caption{Piezoelectric charge density (left) and corresponding piezo-potential in each layer $\ell$ (middle) and at a distance $d_{\rm out}$ above the outer layer (right) across a screw dislocation for a $2N=2,4$ and 6 layer 2H-MoS${}_2$ structure.}
    \label{fig:Piezoelectric_FL}
\end{figure}

In Fig. \ref{fig:Piezoelectric_FL} we show piezoelectric charge density profiles in the vicinity of a dislocation in the middle of a bi-, tetra- and hexalayer of 2H-MoS${}_2$. The three panels on the left column display the on-layer charge densities for each of these films indicating that most of the piezocharge redistribution happens in the two layers near the dislocation plane. We also note that each layer remains overall electrically neutral $\int_{-\infty}^{\infty}d\xi\, \rho^{(\ell)}(\xi) = 0$.

The panels in the middle and in the right of Fig. \ref{fig:Piezoelectric_FL} display the resulting piezo-potential distribution, $\varphi_{\rm FL}$, in each layer and at a distance $d_{\rm out}$ from the outer layer, respectively. The later was obtained by solving Poisson equation in three dimensions (here, we consider a film placed in vacuum),
\begin{equation}\label{PoissonFL}
\begin{split}
    \Bigg[ \nabla^{2} +  4\pi\alpha_{\rm 2D} \sum_{\ell=1}^{2N}& \delta(z-z_{\ell})\nabla_{\rr}^2 \Bigg] \varphi_{\rm FL}(\rr,z) \\
    &= -4\pi \sum_{\ell=1}^{2N}\rho^{(\ell)}(\rr)\delta(z-z_{\ell}).
\end{split}
\end{equation}
Here, we take into account polarization charges $\rho_{\rm ind}^{(\ell)}=\alpha_{\rm 2D}\delta(z-z_{\ell})\nabla_{\rr}^{2}\varphi_{\rm FL}(\rr,z)$, where $\alpha_{\rm 2D}\approx 6.6$ ${\rm \AA}$ is an in-plane polarizability of each MoS${}_2$ monolayer \cite{berkelbach2013theory}.

The Green function for this problem, $G(\qq;z,z')$, can be found using the Fourier-transformed Eq. \eqref{PoissonFL},
\begin{widetext}
\begin{equation}
    \partial_{z}^{2}G(\qq;z,z') - q^{2}G(\qq;z,z') -4\pi\alpha_{\rm 2D} q^{2} \sum_{\ell=1}^{2N} \delta(z-z_\ell)G(\qq;z,z') = \delta(z-z'),
\end{equation}
\end{widetext}
and we use it to compute the piezo-potentials, as
\begin{equation}
    \varphi_{\rm FL}(\rr,z) = -4\pi\int \frac{d^{2}q}{(2\pi)^{2}}\, e^{i\qq\cdot\rr} \sum_{\ell=1}^{2N} G(\qq;z,z_{\ell})\Tilde{\rho}^{(\ell)}(\qq),
\end{equation}
with
\begin{equation*}
    \Tilde{\rho}^{(\ell)}(\qq) = \int d^{2}r\, e^{-i\qq\cdot\rr} \rho^{(\ell)}(\rr).
\end{equation*}

As shown in Fig. \ref{fig:Piezoelectric_FL} (middle panel), a bilayer features the most pronounced potential variations, with a smaller amplitude in thicker films ($N\geq 3$), and potential slowly varying across the layers at each lateral position. In this respect, tetralayers represent an anomaly with much larger potential amplitude in the inner as compared to outer layers. Additionally, we propose that these piezo-potential profiles can be measured through Kelvin probe force microscopy, therefore, in the right panel of Fig. 4, we show the profiles corresponding to 0.5 and 1 nm above the outer layer. For a bilayer, there is a potential jump comparable to the magnitude obtained in each monolayer, while for tetra- and hexalayers the effect of the opposite charges in the neighbouring layers results in an abrupt attenuation of the effects outside the structure. 

In principle, potential profiles displayed in Fig. \ref{fig:Piezoelectric_FL} could trap electrons. However, the analysis of conduction band states of electrons in the vicinity of a dislocation has to take into account the variation of the interlayer hybridisation of the states at the band edges (which is sensitive to the local stacking configuration), leading to the variation of the band energy \cite{ferreira2021band,magorrian2021multifaceted}. The stacking-dependent Hamiltonian for charge carriers in a multilayer system is given by \cite{ferreira2021band}
\begin{widetext}
\begin{equation}\label{N_Ham}
    H_{\Upsilon}^{(2N)} = 
    \begin{pmatrix}
        \varepsilon_{\Upsilon}(\rr_{0}^{(1,2)}) - e\varphi_{\rm FL}^{(1)} & T_{\Upsilon}(\rr_{0}^{(1,2)}) & \cdots & 0 \\
        T_{\Upsilon}^{*}(\rr_{0}^{(1,2)}) & \varepsilon_{\Upsilon}(\rr_{0}^{(1,2)}) + \varepsilon_{\Upsilon}(\rr_{0}^{(2,3)}) - e\varphi_{\rm FL}^{(2)} & \cdots & 0 \\
        \vdots & \vdots & \ddots & \vdots \\
        0 & 0 & \cdots &\varepsilon_{\Upsilon}(\rr_{0}^{(2N-1,2N)}) - e\varphi_{\rm FL}^{(2N)}
    \end{pmatrix},
\end{equation}
where $\Upsilon=Q_{1,2,3}$ for electrons and $\Upsilon=\Gamma$ for holes. Explicit expressions for the stacking dependence of on-layer potential $\varepsilon_{\Upsilon}$, which accounts for spin-orbit splitting for electron bands, and the interlayer coupling $T_{\Upsilon}$ were taken from Ref. \onlinecite{ferreira2021band}.
\end{widetext}

\begin{figure}[h!]
    \centering
    \includegraphics[width=1.0\columnwidth]{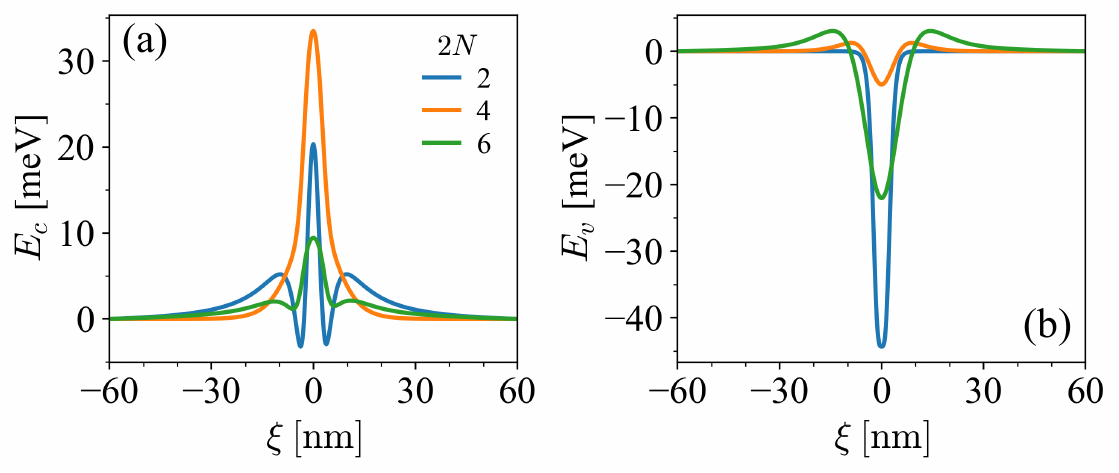}
    \caption{(a) Lowest $Q$ point conduction subband edge modulation and (b) highest $\Gamma$ point valence subband edge modulation across an interlayer screw dislocation in $2N=2,4$ and 6 layer 2H-MoS${}_2$. Energies are calculated with respect to the 2H stacking region.}
    \label{fig:BandEdges}
\end{figure}

In a multilayer film, there would be multiple subbands determined by size quantisation of the electron motion along $z$-axis. To analyse electron states that may be trappe d by the dislocation, we need to focus only on the lowest of those subbands. In Fig. \ref{fig:BandEdges} we plot the resulting effective profiles for (a) the lowest subband energy near the conduction band edge in a MoS${}_2$ film, which appears to be around the $Q$ points, and (b) for the highest subband near the valence band edge, which is at the $\Gamma$ point. Neither of those band edge profiles is able to support a confined state of an electron/hole.

We also find the lack of confinement of the lowest subband electrons/holes in thicker films, up to $2N\sim 30$. However, one may expect a different behaviour of electrons near an isolated screw dislocation in bulk TMD, where the local stacking variation at the dislocation plane is reduced as compared to thin films, so that we analyse this system in the next section.

\section{An isolated perfect screw dislocation in bulk M\lowercase{o}S${}_2$}\label{Sec:Bulk}

\begin{figure*}
    \centering
    \includegraphics[width=1.0\textwidth]{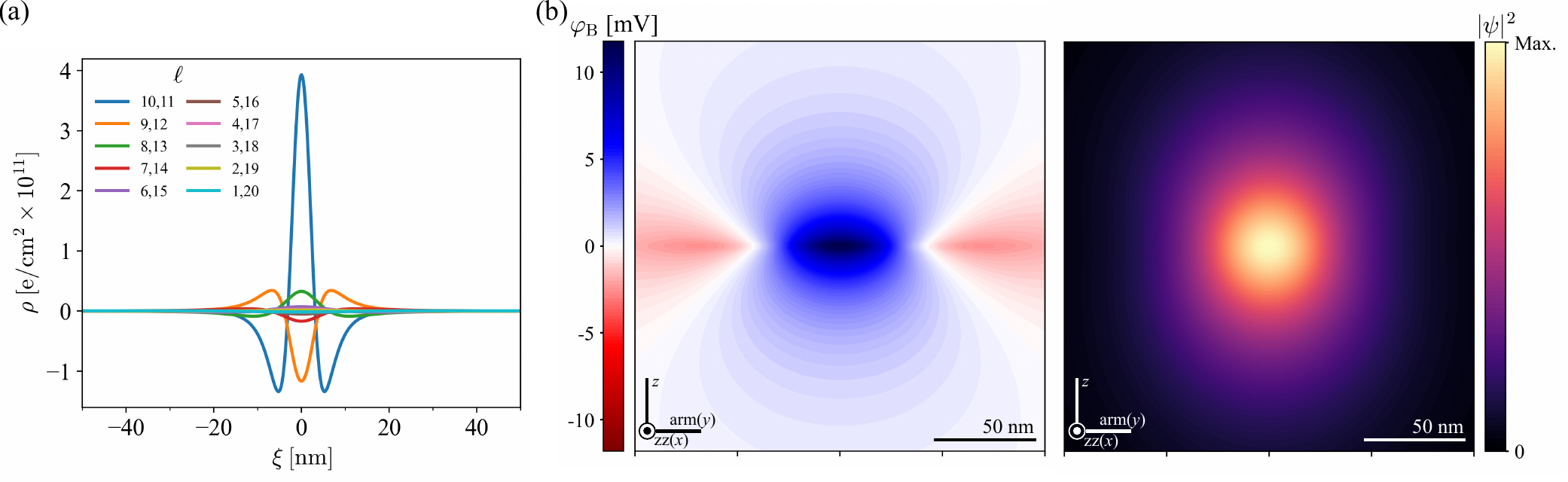}
    \caption{(a) On-layer piezoelectric charge density profiles across a screw dislocation in a 20 layer 2H-MoS${}_2$ film. (b) Piezo-potential in bulk MoS${}_2$ (left) and the corresponding square moduli of the electron bound state (right) for the $Q_{1}$ point.}
    \label{fig:Piezopotential}
\end{figure*}

To cross over from thin-films to bulk crystals, we compute piezoelectric charge profiles for a screw dislocation ($\xi \equiv y$) in a 20 layer MoS${}_2$ film. Those are presented in Fig. \ref{fig:Piezopotential}(a), showing that the non-vanishing contributions to the piezoelectric potential are produced by just five closest layers on each side from the dislocation plane. and that these profiles will be approximately identical in a bulk sample. As a result, we can model the long-range potential distribution in a bulk crystal as determined by a planar distribution of charges summed over those few layers, $\Lambda(\xi)$. Then, the piezo-potential, $\varphi_{\rm B}(\xi,z)$, in the surrounding bulk crystal is described by the solution of a continuum Poisson model,
\begin{equation}
    \big[ \varepsilon_{||}\partial_{\xi}^{2} + \varepsilon_{\perp}\partial_{z}^{2} \big] \varphi_{\rm B}(\xi,z) = -4\pi \Lambda(\xi)\delta(z),
\end{equation}
\begin{equation*}
    \varphi_{\rm B}(\xi,z) = - \int_{-\infty}^{\infty}d\xi'\, \frac{\Lambda(\xi')}{ \sqrt{\varepsilon_{\parallel}\varepsilon_{\perp}}} \ln{\bigg[ \frac{(\xi-\xi')^{2}}{d^{2}} + \frac{\varepsilon_{\parallel}}{\varepsilon_{\perp}}\frac{z^{2}}{d^{2}} \bigg]},
\end{equation*}
where $\varepsilon={\rm diag}(\varepsilon_{\parallel},\varepsilon_{\parallel},\varepsilon_{\perp})$ is a dielectric tensor of bulk MoS${}_2$ ($\varepsilon_{\parallel}=15.9$ and $\varepsilon_{\perp}=6.1$ \cite{laturia2018dielectric,ferreira2022scaleability}). In the integral describing $\varphi_{\rm B}$, $d$ stands for the interlayer distance; however, the exact value of $d$ does not matter due to the overall neutrality of the piezocharge distribution in the dislocation, $\int_{-\infty}^{\infty}d\xi\, \Lambda(\xi) = 0$. 

The resulting profile for this piezo-potential is shown in Fig. \ref{fig:Piezopotential}(b), with a identifiable potential splash near the centre of the dislocation, leading to a 20nm$\times$30nm potential well for electrons (for holes this would be a barrier). For the $Q$ point electrons in bulk MoS${}_2$, the bound states are determined by the Schrodinger equation
\begin{equation}\label{BulkSchr}
    \bigg[ -\frac{\hbar^2}{2}\bigg( \frac{\partial_{\xi}^{2}}{m_{c}^{{\rm arm}}} + \frac{\partial_{z}^{2}}{m_{c}^{z}} \bigg) - e\varphi_{\rm B}(\xi,z)  \bigg] \psi(\xi,z) = \epsilon\psi(\xi,z).
\end{equation}
Here, we distinguish between the $Q_{1}$ point, with effective masses \cite{ruiz2018hybrid}, $m_{c}^{z} = 0.525\, m_{0}$, and $m_{c}^{\rm arm} = 0.735\, m_{0}$, and a pair of $Q_{2}$ and $Q_{3}$ points  $m_{c}^{\rm arm} = 0.587\, m_{0}$. Numerical solution of \eqref{BulkSchr} results in a weakly bound state with energy $\epsilon(Q_{1})=-0.66$ meV for the $Q_{1}$ point (binding of electrons in the other two valleys appears to be even weaker). After comparing binding of electrons by piezo-potential to their mutual repulsion, $V_{\rm C} = e^{2}/(r\sqrt{\varepsilon_{\parallel}\varepsilon_{\perp}})$, we estimate the maximum carrier density that can be trapped by a single dislocation at 
low temperatures as 4.8 $e/\mu$m \footnote{Similarly, electrons in the $Q_{2,3}$ points have bound state energies of $\epsilon(Q_{2,3})=-0.51$ meV, producing a charge carrier density of 3.85 $e/\mu$m.}. Moreover, the estimated small binding energies suggest that all trapped electrons would be evaporated from the dislocation at any reasonable temperature, in particular, at room temperature.

\section{Discussion \& conclusions}\label{Sec:Conclusions}

We have presented a mesoscale model for in- and out-of-plane deformations in the vicinity of an interlayer dislocations in a multilayer MoS${}_2$ film, as well as bulk van der Waals crystal. We find that strain profiles within layers close to the dislocation plane are the same for films various thicknesses (almost the same as for a dislocation in a bilayer \cite{alden2013strain,lebedeva2016dislocations,enaldiev2020stacking,enaldiev2024dislocations}), for the outer layers a weak strain field spread out broader within the layers away from the core of the dislocation, as described by Eq. \eqref{FitOut}. We also calculated the dislocation energy which depends logarithmically on the film thickness, in line with classical dislocation theory \cite{anderson2017theory}.

We also analysed the interplay between potential of piezoelectric charges induced by shear deformations in the layers and the variation of interlayer hybridization across dislocations, to find the band edge profiles created by the dislocation for electrons and holes in MoS${}_2$. Based on this analysis, we find that interlayer screw dislocations in MoS$_2$ are incapable of localising charge carriers. We find that the band edge variation around the dislocation in a thin film does not even provide a trap for either for electrons or holes. We now that weak traps develop only in films with thicknesses of more than 30 layers, however, the estimated binding energies do not exceed 1 meV even in bulk crystals, indicating that such dislocations are not capable of trapping carriers at any practical temperature for device operation, in particular, room temperature.

\section*{Acknowledgements}

I.S. acknowledges financial support from the University of Manchester's Dean's Doctoral Scholarship. This work was supported by EPSRC Grants EP/S030719/1 and EP/V007033/1, and the Lloyd Register Foundation Nanotechnology Grant.

\bibliography{references}

\end{document}